\documentclass[aps,prb,twocolumn,fleqn,showpacs,showkeys]{revtex4}
\usepackage{graphicx}
\usepackage{amssymb}
\usepackage{amsmath}
\usepackage{bm}

\def\da     {\downarrow}
\def\ua     {\uparrow}
\def\rg     {\rangle}

\begin{document}
\title[]{Strong Coupling of a Cavity QED Architecture for a Current-biased Flux Qubit}
\author{Mun Dae \surname{Kim}}
\email{mdkim@yonsei.ac.kr}
\thanks{Fax: +82-2-392-1592}
\author{K. \surname{Moon}}
\affiliation{Institute of Physics and Applied Physics, Yonsei
University, Seoul 120-749, Korea}

\date[]{Received 8 March 2011}

\begin{abstract}
We propose a scheme for a cavity quantum electrodynamics (QED)
architecture for a current-biased superconducting flux qubit with
three Josephson junctions. The qubit operation is performed by
using a bias current coming from the current mode of the circuit
resonator. If the phase differences of junctions are to be coupled
with the bias current, the Josephson junctions should be arranged
in an asymmetric way in the qubit loop. Our  QED scheme provides a
strong coupling between the flux qubit and the transmission line
resonator of the circuit.
\end{abstract}



\maketitle

\section{Introduction}
Superconducting qubits coupled to a quantum mechanical
harmonic oscillator have reached the strong coupling region, where
the coupling strength is much larger than the decay rates of
the cavity and the qubit.
The circuit quantum electrodynamics (QED) scheme has been
applied to the superconducting charge qubit \cite{Blais,Blais07},
phase qubit \cite{Sillanpaa}, and flux qubit \cite{Abd,Yang,Niem,Fedorov}.
%
%
Recently, in Refs. \cite{Abd,Niem,Fedorov} the phase-biasing scheme was studied
to provide a strong coupling strength between the resonator and the flux qubit
by sharing the qubit loop with the resonator.
This galvanic coupling, however, has difficulty in
switching on and off the coupling between the qubit and the resonator,
which is essential for a scalable design.
In this study, by introducing a {\it current-biasing} scheme for the flux qubit,
we offer a new circuit QED architecture,
where the flux qubit is coupled with the {\it current mode of
the resonator}, to obtain a strong coupling between the flux qubit and the resonator
without the galvanic link.
In this scheme, the flux qubit is biased by the oscillating current mode of
the resonator.

The current-biased dc-SQUID qubit (phase qubit) \cite{Martinis,Berkley,Steffen}
is controlled by an electric field, which provides a fast operation.
On the other hand, the flux qubit \cite{Mooij,Niskanen}
is operated at the optimal point where the first-order phase fluctuations vanish.
The present current-biased flux qubit operation is also performed
at a point optimally biased with respect to both the bias current
and the external magnetic flux, which may provide a long coherence time.
%
An oscillating bias current induces a Rabi oscillation between
qubit states.
If the bias current operation of the flux qubit is to be performed,
the number of the Josephson junctions in the flux qubit loop should be
three (in general, an odd number), and the junctions should be arranged in an
asymmetrical way.

\section{Current-biased flux qubit}
\begin{figure}[t]
\vspace{-0cm}
\includegraphics[width=0.45\textwidth]{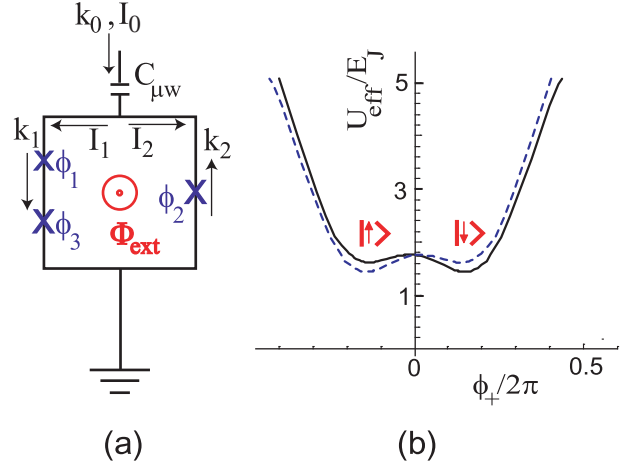}
\vspace{-0cm}
\caption{(Color online) (a) Current-biased flux qubit. Two current states of the
three-Josephson-junction loop interact with the bias currents.
$I_{0}$ is the bias current applied across the capacitance
$C_{\mu\!\rm w}$, and $k_i$'s are the wave vectors of the Cooper pair wave function.
(b) Plot for the effective potential $U_{\rm eff}(\{\phi_\pm\})$
with $f=0.5$ and $\phi_-$=0. The potential tilts due to a finite bias current
$I_0$. The solid (dashed) line corresponds to positive (negative) $I_0$.
}
\label{fig1}
\end{figure}

Figure \ref{fig1}(a) shows the  current-biased flux qubit, where
the current flows through the three-Josephson-junction qubit loop.
In the circuit QED architecture with the flux qubit, which we will
discuss later, the bias current comes from the oscillating current
mode of the resonator. For the time being, we consider an externally applied
oscillating bias current. The Hamiltonian of this system can be
derived semiclassically by using the quantum Kirchhoff relation.

First of all, let's consider a superconducting loop without a Josephson junction,
where the usual fluxoid quantization condition reads  \cite{Tinkham}
\begin{eqnarray}
\label{Tc}
-\Phi_t+(m_c/q_c)\oint \vec{v}_c \cdot d\vec{l}=n\Phi_0,
\end{eqnarray}
with $\vec{v}_c$ being the average velocity of Cooper pairs, $q_c$ being $2e$, and $m_c$ being $2m_e$.
The total magnetic flux threading the loop $\Phi_t$ is the sum of the external and the induced flux
$\Phi_{\rm t}=\Phi_{\rm ext}+\Phi_{\rm ind}$ with  the superconducting unit flux quantum $\Phi_0=h/2e$.
Then, the condition of Eq. (\ref{Tc}) is written as
\begin{eqnarray}
k{\it l}=2\pi (n+ f_t),
\end{eqnarray}
where  $f_t\equiv \Phi_t/\Phi_0=f+f_{\rm ind}$ with $f=\Phi_{\rm ext}/\Phi_0$ and
$f_{\rm ind}=\Phi_{\rm ind}/\Phi_0$, $l$ is the circumference of the loop, and $k$ is the
wave vector of the Cooper pair wavefunction.

For the current-biased flux qubit in Fig. \ref{fig1}(a), the fluxoid quantization
condition in Eq. (\ref{Tc}) is changed due to the phase differences $\phi_i$'s
in the circumference of the qubit loop as follows:
\begin{eqnarray}
\label{fqc}
(k_1+k_2)\frac{l}{2}=2\pi(n+f_t)-\phi_1-\phi_2-\phi_3.
\end{eqnarray}
Usually, for flux qubits, the contribution $(k_1+k_2)l/2$
is negligible in Eq. (\ref{fqc}) which then can be reduced to  $2\pi(n+f_t)-\phi_1-\phi_2-\phi_3=0$.
Rigorously, however, we keep this term for the time being
to derive the Hamiltonian of our qubit system.

The induced flux $\Phi_{\rm ind}$ is given by $\Phi_{\rm ind}=L_s (I_1-I_2)/2$ with
the self inductance $L_s$, and the current $I_{1(2)}$ in Fig. \ref{fig1}(a) is
\begin{eqnarray}
\label{I}
I_{1(2)}=\mp(n_cAq_c/m_c)\hbar k_{1(2)},
\end{eqnarray}
with $n_c$ being the Cooper pair
density and $A$ the cross section of the superconducting loop. Then, by
using Eq. (\ref{I}) and $\Phi_0=h/q_c$,
$f_{\rm ind}=\Phi_{\rm ind}/\Phi_0$ can be represented as $f_{\rm ind}=
-(L_s/L_K)(l/2)(k_1+k_2)/2\pi$ with $L_K = m_cl/An_cq^2_c$ being the
kinetic inductance \cite{flux}.
With this relation the condition in Eq. (\ref{fqc}) becomes
\begin{eqnarray}
\label{fqc2}
\left(1+\frac{L_s}{L_K}\right)(k_1+k_2)\frac{l}{2}
=2\pi\left(n+f-\frac{1}{2\pi}\sum^3_{i=1}\phi_i\right).
\end{eqnarray}

The dynamics of the Josephson junction is described in the
capacitively-shunted model, where the current relation is given by
$I=-I_{\rm c}\sin\phi+C\dot{V}$ with the capacitance of junction
$C$, $\dot{V}=dV/dt$, and voltage $V$ across the junction. This
relation can be rewritten by using the Josephson voltage-phase
relation $V=-(\Phi_0/2\pi){\dot \phi}$ as
\begin{eqnarray}
\label{Jc}
-(n_cAq_c/m_c)\hbar k_i=-I_{{\rm
c}i}\sin\phi_{i}-C_{i}(\Phi_0/2\pi)\ddot{\phi}_{i},
\end{eqnarray}
with the critical current of Josephson junction $I_{{\rm c}i}=2\pi E_{Ji}/\Phi_0$ ($i=1,2,3$)
and the Josephson coupling energy $E_{Ji}$.
Then, from Eqs. (\ref{fqc2}) and (\ref{Jc}) with
$k_0=k_1-k_2$, we have the quantum Kirchhoff relation
\begin{eqnarray}
\label{motion}
\left(\frac{\Phi_0}{2\pi}\right)^2C_{i}\ddot{\phi}_{i}&=&
\frac{\Phi^2_0}{2(L_s+L_K)\pi}\left(n+f-\frac{1}{2\pi}\sum^3_{i=1}\phi_i\right)\nonumber\\
&&-E_{Ji}\sin\phi_{i}\mp\frac{\Phi_0I_0}{4\pi},
\end{eqnarray}
where $I_{0}=-(n_cAq_c/m_c)\hbar k_{0}$.
Here, for $i=1,3$,  the last term of Eq. (\ref{motion}) is $+\Phi_0I_0/4\pi$
whereas for $i=2$ the sign is reversed as $-\Phi_0I_0/4\pi$.
For usual flux qubits, $L_K/L_s \sim 0.01$ \cite{Mooij};
thus, hereafter $L_K$ is neglected for simplicity.

The equation of motion, Eq. (\ref{motion}), can be obtained
from the Lagrange equation $(d/dt)\partial {\cal L}/\partial \dot{\phi}_i-\partial {\cal L}/\partial \phi_i=0$
with the Lagrangian
\begin{eqnarray}
\label{Lag}
{\cal L}(\phi_i,\dot{\phi}_i)&=&\sum^3_{i=1}\frac12 C_i\left(\frac{\Phi_0}{2\pi}\right)^2\dot{\phi}^2_i
-U_{\rm eff}(\{\phi_i\}),\\
\label{Ueff}
U_{\rm eff}(\{\phi_i\})&=&\sum^3_{i=1}E_{Ji}(1-\cos\phi_i)+\frac{\Phi_0I_0}{4\pi}(\phi_1+\phi_3-\phi_2)\nonumber\\
&&+\frac{\Phi^2_0}{2L_s}\left(n+f-\frac{1}{2\pi}\sum^3_{i=1}\phi_i\right)^2,
\end{eqnarray}
where the first term in Eq. (\ref{Lag}) comes from the charging energy
of the qubit system, $E_C=
\sum^3_{i=1}Q^2_i/2C_i$ with $Q_i=C_i(\Phi_0/2\pi)\dot{\phi_i}$.
%
The second term of Eq. (\ref{Ueff}) has a finite value owing to
the asymmetry of the qubit loop, giving rise to the coupling between
the bias current and the flux qubit.

In the usual experiments for flux qubits,
$\Phi^2_0/2L_s \sim O(10^3E_J)$ is much larger than the other energy
scale in the Lagrangian of Eq. (\ref{Lag}). Hence, the last term in
Eq. (\ref{Ueff}) can be removed, leaving  the  constraint
$g(\phi_i)=\phi_1+\phi_2+\phi_3-2\pi(n+f)=0$,
which is  the familiar flux quantization condition.
The Lagrangian then has the effective potential
\begin{eqnarray}
U_{\rm eff}(\{\phi_i\})\!=\!\!\!\sum^3_{i=1}E_{Ji}(1\!-\!\cos\phi_i)
\!+\!\!\frac{\Phi_0I_0}{4\pi}(\phi_1\!+\!\phi_3\!-\!\phi_2)
\end{eqnarray}
with the constraint $g(\phi_i)=0$. The Lagrange
equation of motion with the above constraint
produces the Kirchhoff
equations in the qubit loop of Fig. \ref{fig1}(a): $I_0=I_1+I_2$ and $I_1=I_3$.

\begin{figure}[t]
\vspace{0.5cm}
\includegraphics[width=0.4\textwidth]{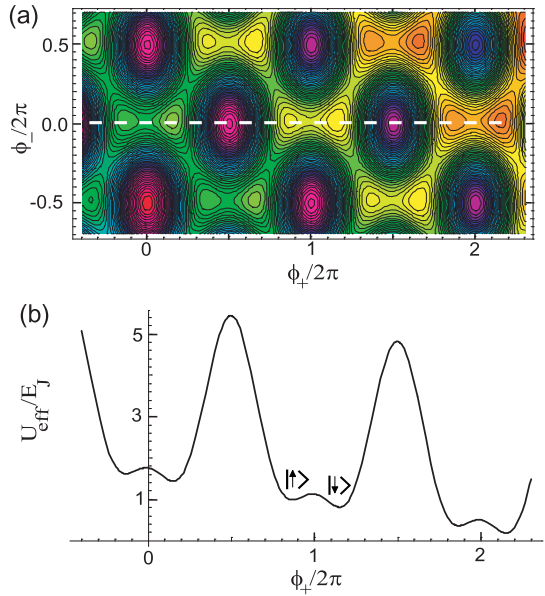}
\vspace{0.cm}
\caption{(Color online) (a) Effective potential of the current-biased flux qubit
in the plane of $(\phi_+, \phi_-)$, where the effective potential of the local minima decreases as
$\phi_+$ or $\phi_-$ increases. Here, we set $I_0/I_{\rm c}$=0.05, $f=0.5$ and $E_{J1}/E_J$=0.8.
(b) Plot for $U_{\rm eff}(\{\phi_\pm\})$
along the dashed line in the upper panel.
}
\label{fig2}
\end{figure}

We  introduce a coordinate transformation such as
$\phi_+=(\phi_2+\phi_3)/2$ and $\phi_- = (\phi_2-\phi_3)/2$. In the
usual flux qubit experiment, the two Josephson junctions are nearly
identical; thus, we set $E_{J2}=E_{J3}=E_J$. Although one can
treat the general case numerically, this case gives a clear and intuitive
picture for our qubit system. In this case, the effective potential
is given by
\begin{eqnarray}
\label{UeffT}
U_{\rm eff}(\{\phi_\pm\})&=&-E_{J1}\cos(2\pi f-2\phi_+)
-2E_{J}\cos\phi_+\cos\phi_- \nonumber\\
&&+\frac{\Phi_0I_0}{2\pi}(\pi f-\phi_+ -\phi_-).
\end{eqnarray}
Without the last term representing the linear coupling of the phase
to the external bias current $I_0$, the effective potential of Eq. (\ref{UeffT}) can have
local minima only for $\cos\phi_-=\pm 1$, {\it i.e.}, $\phi_-= j\pi$,
with $j$ being an integer, where the qubit states are formed (See Fig. \ref{fig2}(a)).

Let the flux qubit be at the degeneracy point $f=0.5$.
When there is no bias current $I_0=0$, $\phi_+ \approx
\pi/3(-\pi/3)$ for the counterclockwise (clockwise) current state
$|\downarrow\rangle ~(|\uparrow\rangle)$ with $\phi_-= 0$,
and  the qubit energy levels corresponding to the local minima of $U_{\rm eff}(\{\phi_\pm\})$
are degenerate,  $E_{\downarrow}=E_{\uparrow}$.
In this case, the qubit is optimally biased
with respect to both the current $I_0$ and the  magnetic flux $f$.
A finite bias current tilts the effective potential as shown in Fig. \ref{fig2}(b),
where  the direction of energy level tilt depends on the sign of $I_0$.
Consequently, the effective potential of Eq. (\ref{UeffT}) is expressed as
$(E_{\downarrow}-\Phi_0I_0\alpha/2\pi)|\downarrow\rangle\langle\downarrow|
+ (E_{\uparrow}+\Phi_0I_0 \alpha/2\pi)|\uparrow\rangle\langle\uparrow|$ apart from the constant,
where
\begin{eqnarray}
\alpha \approx |\phi_+|\approx \pi/3.
\end{eqnarray}

The transitions between the two states,
$|\downarrow\rangle$ and $|\uparrow\rangle$, are induced by the charging
energy given by the first term of the Lagrangian in Eq. (\ref{Lag}).
Using the tight-binding approximation, we can write the Hamiltonian for our qubit system as
\begin{eqnarray}
\label{H}
{\cal H} \!&=&\! E_{\downarrow}|\downarrow\rangle\langle\downarrow|\!+\!
E_{\uparrow}|\uparrow\rangle\langle\uparrow|
- t_q(|\downarrow\rangle\langle\uparrow|\!+\!|\uparrow\rangle\langle\downarrow|) \nonumber\\
&& -\frac{\Phi_0I_0}{2\pi}\alpha (|\downarrow\rangle\langle\downarrow| -
|\uparrow\rangle\langle\uparrow|),
\end{eqnarray}
where $t_q$ is the transition rate between
$|\downarrow\rangle$ and $|\uparrow\rangle$ states.

In order to operate the single qubit states,
we apply a oscillating current,
\begin{eqnarray}
I_0=-I_{\rm b}\sin\omega t.
\end{eqnarray}
As shown in Fig. \ref{fig1}(b), the effective potential vibrates with the frequency $\omega$,
which produces oscillating diagonal terms in Eq. (\ref{H}).
In transformed coordinates, these terms appear off-diagonal,
giving rise to a Rabi oscillation.
The qubit states, $|0\rg$ and $|1\rg$, are the symmetric and the antisymmetric superpositions
of $|\da\rg$ and $|\ua\rg$, such that
\begin{eqnarray}
|0\rg&=&(|\da\rg+|\ua\rg)/\sqrt{2},\nonumber\\
|1\rg&=&(|\da\rg-|\ua\rg)/\sqrt{2}.
\end{eqnarray}
In the basis of $\{|0\rangle, |1\rangle\}$, the Hamiltonian is represented as
\begin{eqnarray}
\tilde{\cal H}=\frac{\hbar\omega_0}{2}\sigma_z +g\sin\omega t
\sigma_x, \label{OneT}
\end{eqnarray}
where $\hbar \omega_0=2t_q$ is the qubit frequency, and $\sigma_{x,z}$ and $I$
are the Pauli  and the identity matrices, respectively.
The coupling strength $g$ between the oscillating current and the qubit is given by
\begin{eqnarray}
\label{g}
g\equiv \frac{\Phi_0I_{\rm b}}{2\pi}\alpha.
\end{eqnarray}
Near resonance, $\omega\approx \omega_0$, with
a weak coupling $g/\hbar \ll \omega_0$,
one can apply the rotating wave approximation. \cite{Jaynes}
Then, we can observe a Rabi oscillation between the two qubit states $|0\rangle$ and$|1\rangle$
with a Rabi frequency $\Omega_{\rm R}^0=g/\hbar$.


Although we do not present it explicitly, in general the number of Josephson junctions
can be any odd integer larger than one.
In this case, at the two sides of the qubit loop divided by the bias current line,
the numbers of Josephson junctions should be different from each other so that
the symmetry of the flux qubit loop is broken.
Then, the resulting coupling  between the current and the phase of junction
enables current-driven operation of the flux qubit.
%
For the dc-SQUID (2-junction) qubit, the bias current flows across the Josephson junctions,
but the symmetry of the dc-SQUID does  not allow  bias-current
operation of the qubit states.


\section{Circuit QED}
\begin{figure}[t]
\vspace{0.5cm}
\includegraphics[width=0.48\textwidth]{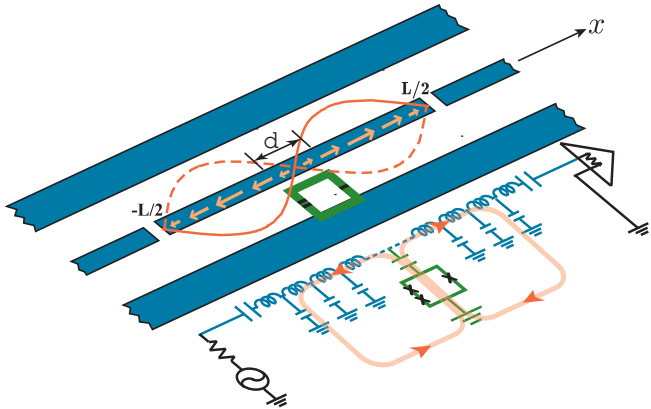}
\vspace{0.3cm}
\caption{(Color online) Schematic diagram  and equivalent lumped circuit representation
of a three-junction flux qubit coupled to a transmission line resonator.
The flux qubit interacts with the current mode of the resonator. The arrows
in the schematic diagram and equivalent circuit show the flow of oscillating current.
As shown in the equivalent circuit, the oscillating current flows between the resonator and the
qubit through the capacitance.
Here, $L$ is the length of the resonator, and $d$ is the width of qubit loop.}
\label{fig3}
\end{figure}

The present current-biasing scheme for the flux qubit is implemented in
the circuit QED architecture. In Fig. \ref{fig3}, the qubit is
coupled to the resonator by an ac current flowing through a
capacitance. The Lagrangian ${\cal L}_r$ of the transmission line
resonator consists of the charge density mode $ q(x,t)$ and the
current density mode $I(x,t)$:
\begin{eqnarray}
{\cal L}_r=\int^{\frac{L}{2}}_{-\frac{L}{2}}dx\left(\frac{l}{2}I^2(x,t)-\frac{1}{2c} q^2(x,t)\right),
\end{eqnarray}
where $l$ and $c$ represent the inductance and the capacitance
per unit length, and $L$ the length of the resonator.
Introducing the variable $\theta(x,t)\equiv \int^x_{-L/2}dx'q(x',t)$,
the Lagrangian becomes
\begin{eqnarray}
{\cal L}_r=\int^{L/2}_{-L/2}dx[(l/2)(\partial_t\theta)^2-(1/2c) (\partial_x \theta)^2],
\end{eqnarray}
where the voltage and the current  of the resonator are given by
$V(x,t)=\frac{1}{c}\frac{\partial\theta(x,t)}{\partial x}$ and
$I(x,t)=\frac{\partial\theta(x,t)}{\partial t}$, respectively.
For example, the voltage and the current of the resonator for the second harmonic mode can be represented in terms of
the boson creation and annihilation operators $a^\dagger(t)$ and $a(t)$ as
 \cite{Blais}
\begin{eqnarray}
\label{VI}
V(x,t)&=&\sqrt{\frac{\hbar\omega_r}{cL}}\cos\left(\frac{2\pi x}{L}\right)[a(t)+a^\dagger(t)],\nonumber\\
I(x,t)&=&-i\sqrt{\frac{\hbar\omega_r}{lL}}\sin\left(\frac{2\pi x}{L}\right)[a(t)-a^\dagger(t)],
\end{eqnarray}
where $\omega_r=2\pi v/L$ and $v=\sqrt{1/lc}$.
The current profile for the second harmonic mode is shown in Fig. \ref{fig3}.

In the circuit QED scheme with superconducting charge qubit,
the qubit interacts with the voltage mode $V(x,t)$. In contrast, the present current-biased flux
qubit is coupled to the temporal fluctuation of local charge density, $I(x,t)$,
corresponding to the bias current applied to the qubit.
In Fig. \ref{fig3} the resonator and the qubit are coupled by a capacitance
through the region $-d/2<x<d/2$, where the charge fluctuation ${\dot q}(x,t)$ in the resonator
produces the current flow into the qubit,
$I_0(t)=\int^{d/2}_{-d/2}{\dot q}(x,t)dx.$
As a result, following the current conservation condition in the resonator,
${\dot q}(x,t)=\partial I(x,t)/\partial x$, and  Eq. (\ref{VI}), the current flowing into
the qubit with width $d$ is given by
\begin{eqnarray}
\label{I0}
I_0(t)&=&I(d/2,t)-I(-d/2,t)\nonumber\\
&=&-2i\sqrt{\frac{\hbar\omega_r}{lL}}\sin\left(\frac{\pi d}{L}\right)[a(t)-a^\dagger(t)].
\end{eqnarray}
The interaction between the flux qubit and the bias current $I_0(t)$
of the resonator is given by the last term of Eq. (\ref{H}).
Then, from Eqs. (\ref{H}) and (\ref{I0}), the total Hamiltonian
at the degeneracy point ($E_\downarrow=E_\uparrow$) is given by
\begin{eqnarray}
 {\cal H}&=&\hbar\omega_r\left(a^\dagger
a+\frac12\right)-t_q(|\downarrow\rangle\langle\uparrow|\!+\!|\uparrow\rangle\langle\downarrow|)\nonumber\\
&&+i g(|\downarrow\rangle\langle\downarrow| -
|\uparrow\rangle\langle\uparrow|)(a-a^\dagger),
\end{eqnarray}
where the first term comes from the oscillating mode in the
resonator. The first two terms of Eq. (\ref{H}) disappear
at the degeneracy point.
In the basis of $|0\rangle$ and $|1\rangle$, the Hamiltonian is
written as
\begin{eqnarray}
\label{Hc} {\tilde {\cal H}}=\hbar\omega_r a^\dagger a
+\frac{\omega_a}{2}\sigma_z+i g\sigma_x(a-a^\dagger),
\end{eqnarray}
with $\hbar \omega_a=2t_q$.
Here, the single qubit gate is performed by a resonant
external driving microwave field \cite{Blais07}.

Since usually $d\ll L$, we have the expression of the coupling $g$ as follows:
\begin{equation}
\label{gc}
g \approx \alpha \Phi_0 \sqrt{\frac{\hbar\omega_r}{lL}}\left(\frac{d}{L}\right).
\end{equation}
Using the usual experimental values for the parameters \cite{Abd} such that
$d=3 ~\mu$m, $L$=5 mm, $lL$=2.5 nH, $\omega_r/2\pi=$15 GHz, and $\alpha=\pi/3$,
we estimate the coupling strength to be $g/h\sim 120$ MHz.
For inductive coupling between the qubit and the resonator,
the coupling strength, $g_{\rm IC}=\Phi I_c\sin\alpha$
with $\Phi$ being the magnetic flux threading the qubit due to
the resonator current, can also be estimated to be
\begin{eqnarray}
g_{\rm IC}\approx \frac{\mu_0 d^2}{2\pi R}\sqrt{\frac{\hbar\omega_r}{lL}}I_c\sin\alpha,
\end{eqnarray}
where $R$ is the mean distance  between the resonator and the qubit loop
and $I_c=2\pi E_J/\Phi_0$.
With the same parameter values, we see that $g_{\rm IC}$ is smaller than $g$
by an order of magnitude.
We obtain the inductive coupling strength $g_{\rm IC}/h\sim 12$ MHz with
$R=5~\mu$m and $E_J/h=200$ GHz.
Note that since the qubit is located at the nodal point of the current oscillation for the second harmonic mode,
the inductive coupling at this point is negligible.
The above value of inductive coupling has been obtained for the first harmonic mode.
We have also assumed that the capacitance density is nearly uniform along the resonator.
In reality, the capacitance around the center of the resonator can be much higher due to
the presence of the qubit loop, which can potentially  provide a much
stronger coupling $g$ in the present scheme.

\section{Decoherence property}

The qubit state of the present current-biased flux qubit is
driven in a different way from that of the flux-driven flux qubit;
thus, the decoherence property will also differ from each
other. According to a recent review for the phase qubit
\cite{QIP}, the dephasing times of the phase qubit and the
three-Josephson junction flux qubit are comparable with each
other. Since the only difference between the present
current-biased flux qubit and the usual current-biased dc-SQUID
qubit (phase qubit) is the number of Josephson junctions in the
qubit loop, the decoherence property of both qubits can be analyzed
in a similar way.


In a recent experiment \cite{Steffen2} for the phase qubit, the
capacitance of Josephson junction is $\sim$ 50fF while the
three-Josephson junction flux qubit has a typically smaller value of
capacitance of $\sim$ 3fF ~\cite{Wal} with a small area of the
Josephson junction. For both qubits, a large shunted capacitance
may reduce the decoherence from charge fluctuation. We employ the
typical parameters of the three-Josephson junction flux qubit for
the present current-biased qubit. Since both the capacitance and
the critical current of the Josephson junction scale as the area of
the junction, the critical current $I_c$ of junction is also much
smaller.

For the present current-biased flux qubit, the dephasing due to the
bias-current noise may cause  decoherence of the qubit state,
as in the phase qubit \cite{QIP}.  The current noise is related to
the $1/f$ charge noise with a spectral density $S^*_q(1 {\rm Hz})/f$.
The phase noise is given by $\langle\phi^2\rangle\approx
[S^*_q(1{\rm Hz})/C\Delta U] (2/3)\ln(1.778\omega_{10}t)$,
\cite{Nam} where $C$ is the capacitance of the Josephson junction,
$\Delta U$ is the barrier height of the potential, and
$\omega_{10}$ is the qubit level spacing. Although for the present
qubit,  the spectral density $S^*_q(1 {\rm Hz})$ and the
capacitance $C$ are small, these contributions cancel each other in
$\langle\phi^2\rangle$ because both the spectral density and the
capacitance scale as the area of the junction \cite{Nam}. Further,
the qubit level spacing $\omega_{01}$ is similar between the two
qubits. For the operation of the present qubit,
we need not tilt the potential; thus, the barrier hight of the
potential  $\Delta U$ remains large. Hence, we consider
the dephasing due to current noise in present qubit to still
be comparable to the flux-driven flux qubit.
On the other hand, the phase noise due to the critical-current fluctuation and
to the flux $1/f$ fluctuations is given by $\langle\phi^2\rangle\approx (S^*_I(1{\rm
Hz})L_J/\Delta U) \ln(0.401/f_m t)(\omega_{01}t)^2/6$  \cite{Nam},
with the Josephson inductance $L_J=\Phi_0/2\pi I_c\cos\phi$ and
the low-frequency cutoff $f_m$. For the phase noise,
the argument is similar because
both the spectral density $S^*_I(1{\rm Hz})$ and the critical
current $I_c$ scale as the area of the junction.

\section{Summary}
We propose a new circuit QED architecture for the three-Josephson-junction flux qubit,
where the flux qubit is coupled to the temporal charge density fluctuations of the
transmission line resonator.
The flux qubit is controlled by using a bias current.
When three Josephson junctions are arranged asymmetrically,
the energy levels of  two qubit states with different
chiralities couple to the external bias current.
Rabi oscillations can be induced by an ac current at an optimal point
with respect to both the bias current and the external magnetic flux.
Remarkably, the coupling between
the qubit and the resonator is strongly enhanced compared to conventional inductive coupling.

\begin{acknowledgments}
We wish to thank S. Girvin for useful discussions and valuable suggestions.
This work was supported by the Korea Research Foundation Grant
funded by the Korean Government (MOEHRD, Basic Research Promotion
Fund) through KRF-2008-313-C00243.
\end{acknowledgments}

\end{document}